\documentclass[conference]{IEEEtran}
\IEEEoverridecommandlockouts
\usepackage{cite}
\usepackage{amsmath,amssymb,amsfonts}
\usepackage{algorithmic}
\usepackage{graphicx}
\usepackage{textcomp}
\usepackage{xcolor}
\usepackage[utf8]{inputenc}
\usepackage{colortbl}
\usepackage{url}
\usepackage{microtype}
\usepackage{multirow}
\usepackage{booktabs}
\usepackage{tcolorbox}

\newcommand{\CalloutFigure}[1]{Fig.~\ref{#1}}

\newcommand{\CalloutTable}[1]{Table~\ref{#1}}

\newcommand{\be}{\begin{enumerate}}
\newcommand{\ee}{\end{enumerate}}
\newcommand{\bi}{\begin{itemize}}
\newcommand{\ei}{\end{itemize}}

\begin{document}

\title{OpenCBS: An Open-Source COBOL Defects Benchmark Suite}

\author{\IEEEauthorblockN{Dylan Lee\IEEEauthorrefmark{1}, Austin Henley\IEEEauthorrefmark{2}, Bill Hinshaw\IEEEauthorrefmark{1}, and Rahul Pandita\IEEEauthorrefmark{1}\IEEEauthorrefmark{3}}
\IEEEauthorblockA{
\textit{\IEEEauthorrefmark{1}Phase Change Software}, Golden, USA\\
\textit{\IEEEauthorrefmark{2}Microsoft}, Redmond, USA\\
\textit{\IEEEauthorrefmark{3}GitHub Inc.}, San Francisco, USA\\
dlee@phasechange.ai, austinhenley@microsoft.com, bhinshaw@phasechange.ai, rahulpandita@github.com}
}

\maketitle

\begin{abstract}

As the current COBOL workforce retires, entry-level developers are left to keep complex legacy systems maintained and operational.
This creates a massive gap in knowledge and ability as companies are having their veteran developers replaced with a new, inexperienced workforce.
Additionally, the lack of COBOL and mainframe technology in the current academic curriculum further increases the learning curve for this new generation of developers.
These issues are becoming even more pressing due to the business-critical nature of these systems, which makes migrating or replacing the mainframe and COBOL anytime soon very unlikely. 
As a result, there is now a huge need for tools and resources to increase new developers' code comprehension and ability to perform routine tasks such as debugging and defect location.
Extensive work has been done in the software engineering field on the creation of such resources.
However, the proprietary nature of COBOL and mainframe systems has restricted the amount of work and the number of open-source tools available for this domain.
To address this issue, our work leverages the publicly available technical forum data to build an open-source collection of COBOL programs embodying issues/defects faced by COBOL developers.
These programs were reconstructed and organized in a benchmark suite to facilitate the testing of developer tools.
Our goal is to provide an open-source COBOL benchmark and testing suite that encourage community contribution and serve as a resource for researchers and tool-smiths in this domain.
\end{abstract}


\section{Introduction}
COBOL and mainframe has provided longstanding, reliable service in business, finance, and healthcare systems.
Because of this essential role, they have withstood the advancement and influx of new technologies, and even today are still in widespread use. 
In fact, the most recent numbers demonstrate the importance of these systems, with \textit{``over 220 billion lines of COBOL code being used in production today, and 1.5 billion lines are written every year''}~\cite{taulli_2020}.
Not only are these systems still in active development but they still handle a huge number of transactions, ``\textit{on a daily basis, COBOL systems handle USD 3 trillion in commerce''}~\cite{cassel_2021}.

Today, as a generation of veteran COBOL developers and mainframe experts are retiring, there is a tremendous need for the recruitment and training of a new workforce. 
Tools to assist with tasks such as code navigation, analysis, debugging, and defect location will be vital to training this influx of new COBOL developers.
However, compared to contemporary technology stacks, the creation of such tools for COBOL and mainframe developers has been relatively scarce.
This is partly due to the potentially sensitive and proprietary nature of business's COBOL codebases, leaving a lack of large-scale open-source COBOL repositories to learn from.
Furthermore, COBOL has slowly been phased out of most academic curricula \cite{carr2003continued, carr2000case} and has received very little attention from scientific and academic researchers. 
For these reasons, a multitude of roadblocks and barriers to entry exist for new COBOL developers.
The creation and study of tools to help overcome these barriers could provide tremendous benefits.

The goal of this work is to provide an open-source test suite made up of \textit{defective} COBOL programs. 
This test suite gives COBOL tool-smiths and researchers a publicly available set of programs that could be utilized for testing, users studies, and other types of empirical research. 
To produce this suite we utilized the vast amount of data contained in COBOL-based forums and message boards by collecting posts containing questions about COBOL issues and errors.
The information in these posts was then evaluated, resulting in the recreation of \textit{defective} COBOL programs. 
Test suites like this provide a much-needed starting point for the building of tools to assist developers in this area. 
We also hope it encourages community contributions and facilitates future research and work on this topic.

Overall, our work makes the following contributions:
\begin{itemize}
 \item An open-source test suite of COBOL programs containing both defects and fixes.~\cite{projectWeb}
 \item A classification of the types of defects present in the testing suite.
 \item A collection of Mainframe/COBOL software development communities.
\end{itemize}
\section{Related Work}
To understand the need for COBOL defect benchmarks, we investigated (1) COBOL in industry and research, (2)  the motivation, creation, and usage of software benchmarks, and (3) classifications for defects.

\subsection{COBOL}
The common business-orientated language, or COBOL, was designed as a programming language for business, finance, and other administrative tasks.
By utilizing a more readable syntax than other languages at the time, COBOL allowed non-technical workers to automate things like managing payroll, budgets, and other business-related data processing.
After its introduction, COBOL's status in the business and financial sector continued to grow, and even today is the code behind some vital systems.
COBOL's legacy status and the fact that it was targeted towards the usability in business applications are factors that make studying its developers unique. 
Most COBOL code-bases have been in existence for decades, with unknown amounts of documentation, and unavailable to the public.
This means very little work has been done to understand how new developers build code comprehension, navigate, write, and debug.
Additionally, many other questions haven't been explored due to the challenges of researching developers in this area.

Despite the original language being over 60 years old, progress is still being made~\cite{de2018cobol, litecky1976study, mossienko2003automated, rodriguez2011bottom, sellink2002restructuring, sneed2001extracting, sneed2010migrating, sneed2013migrating, veerman2006cobol, ciborowska2021contemporary}.
Sneed et al. have worked on both the process for extracting knowledge from COBOL systems~\cite{sneed2001extracting}, as well as multiple works on the process of migrating COBOL systems to Java~\cite{sneed2010migrating, sneed2013migrating}.
De Marco et al.~\cite{de2018cobol} also documents the process for migrating COBOL on the mainframe to Java on Linux for a major newspaper company. 
This work also provides information on the automated generation of Java from COBOL source code, and how the authors tested to ensure functionality equivalence between the modernized and legacy system. 
In another work on migrating COBOL, Rodriguez et al.~\cite{rodriguez2011bottom} present two approaches for migrating from COBOL to a web service-based system and compare and contrast each using a case study on a real-world system.
Similarly and more recently, additional work has been done on migrating COBOL to Java using an automated translating method by Mossienko et al.~\cite{mossienko2003automated}. 
In another context, Sellink et al. \cite{sellink2002restructuring} describe the process of restructuring COBOL code before performing large-scale renovations.  

Additional to work on migrating and modernizing legacy systems, some studies have investigated the defects and errors experienced in COBOL development.
In a study targeted towards new developers learning COBOL, Litecky et al.~\cite{litecky1976study} gives a breakdown of the most common errors experienced by students.
Another work on COBOL defects by Veerman et al.~\cite{veerman2006cobol} builds a "mine-detector" that attempts to detect unintentional programming errors the authors refer to as "mines", which is then evaluated in an industry setting.
Lastly, we found at least one work that provides an example of some type of open-source collection of COBOL programs~\cite{parser}.
However, these example COBOL programs were collected for the purpose of providing a proof of concept for COBOL-based AST generation and don't provide any type of information on COBOL defects or errors. 


\subsection{Benchmark Creation and Usage}
Benchmarks are an essential part of the software development ecosystem.
They allow reliable, reproducible, and comprehensive testing and evaluation of software systems.
Without benchmarks, comparing changes and advancements in system performance is a much more challenging task.

There is a multitude of previous work centered around the creation, evaluation, and usage of benchmark test suites.
Due to the importance of these test suites, some research is dedicated solely to providing recommendations for the creation of benchmark suites. 
Huppler et al.~\cite{huppler2009art, v2015build} define the 5 most important characteristics of a benchmark: relevance, reproducibility, fairness, verifiability, and usability. 
This work also provides recommendations and insight on how best to balance these features to build an effective and beneficial benchmark.

Other work is focused on comparing and evaluating current benchmark and test systems across different domains and software systems.
For example, Ivanov et.al.~\cite{ivanov2015big} gather and compare multiple different benchmarks to provide a compendium for benchmarks in the context of big data.
This provides a framing for the current state of benchmarks available in different domains and provides recommendations for use cases.
Additionally, there is existing work that publishes new benchmark suites, either providing new features or improving the current state of the art~\cite{hara2008chstone, heckman2008establishing, tron2007benchmark, zhou2018fault}. 
Somewhat similar to our work, Heckman et al.~\cite{heckman2008establishing} built a benchmark to evaluate the performance of static analysis methods for fault detection.
Likewise, Zhou et al.~\cite{zhou2018fault} created an open-source micro-service benchmark system by first utilizing a survey to understand the most common faults in micro-services, and then recreating 22 of the most common faults in their benchmark system.
Yet another work, this one from Hara et al. \cite{hara2008chstone}, assembles a benchmark system for C-based high-level synthesis, and the authors cite the importance of benchmark systems for comparison and evaluation of new ideas.
Lastly, Tron et al. \cite{tron2007benchmark} created a benchmark system that the authors then used to perform a novel comparison between different 3-D motion segmentation algorithms.

\textit{Although there are numerous benchmarks, there is a clear lack of this type of work for COBOL and mainframe development.
This is the gap we attempt to address in this work by building off of previous work on the benchmark creation, testing, and evaluation, and then applying these methods to the domain of COBOL development and mainframe technology.}

\subsection{Defect Classification}
Our software defect types we used were directly based on the IBM Orthogonal Defect Classification standards (ODC) \cite{chillarege1992orthogonal, ibmresearch}.
This classification method provides an in-process measurement system for identifying software defects. 
IBM's ODC has become a well-established method for defect analysis and is used in many other works, \cite{bridge1998orthogonal, huang2015autoodc, lopes2020automating, tiejun2008defect}.
For example, Bridge et al. \cite{bridge1998orthogonal} use ODC to measure development progress and identify process problems in an industry setting, demonstrating the feasibility and advantages of applying ODC.
Tiejun et al. \cite{tiejun2008defect} apply ODC in a similar way, to establish a concrete workflow for defect tracing, therefore improving the ability to prevent defects during development.
Huang et al. \cite{huang2015autoodc} attempt to reduce the amount of human-intensive work involved in applying ODC by creating an automated system that takes advantage of multiple machine learning techniques.
Following that same idea, Lopes et al. \cite{lopes2020automating} also apply different machine learning methods, trained on bug reports, to determine the most effective and accurate model for assigning orthogonal defect classifications.
More recently, Cibororowaska et al.~\cite{ciborowska2021contemporary} used ODC defect categories to understand COBOL developer perspectives on defects.
The longstanding and well-established effectiveness of classifying software defects using ODC, which is demonstrated in these works and many others, is what led us to adopt this method.

\textit{Our work builds from this type of prior research and extends it in a new direction by providing an open-source COBOL benchmark based on community Q\&A and built by a professional COBOL developer.}
\section{Methodology}
\begin{figure*}[!ht]
 \centering
 \includegraphics[width=0.85\textwidth]{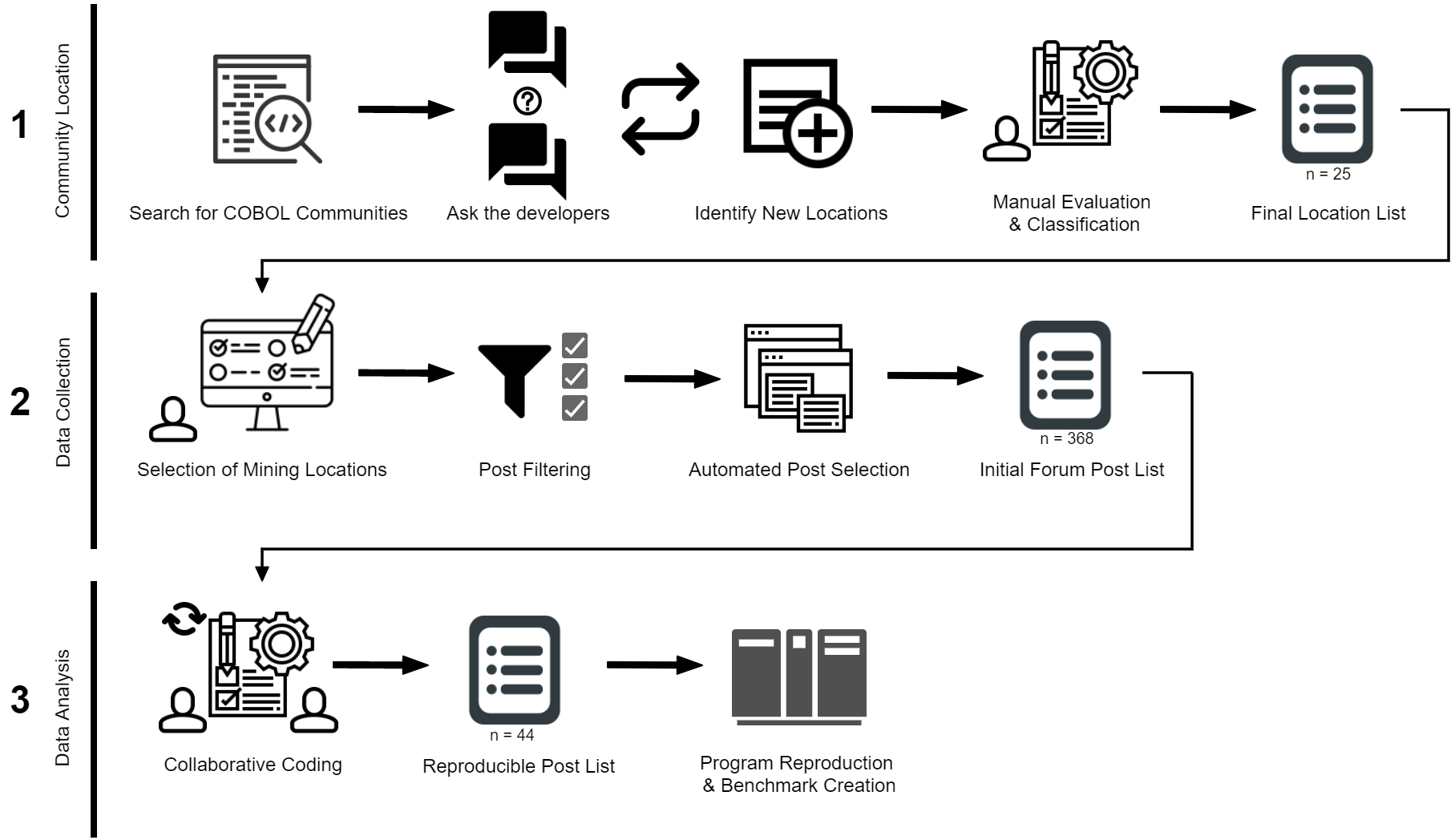}
 \caption{Overview of the methodology used to (1) locate communities, (2) collect data, and (3) analyze data to produce the benchmark suite.}
 \label{fig:method}
\end{figure*}

Our process for collecting and recreating common COBOL defects consisted of three main phases: (1) community identification, (2) data collection, and (3) data analysis.
The community location stage was focused on ensuring as much coverage as possible of all the places technical COBOL-based questions are asked.
In the data collection phase, we evaluated locations for relevance and began collecting and filtering posts containing questions on COBOL defects or errors.
Finally, the collected data was analyzed to determine reproducibility and defect type, resulting in a collection of recreated defective COBOL programs that make up the benchmark suite.
This process is represented visually in Figure~\ref{fig:method}, where each stage consists of multiple steps, which are detailed in the following sections.

\subsection{Community Locations}
The first major stage of our process was to find and identify the locations that COBOL developers use to troubleshoot and seek assistance.
The goal of this search was to find places in which developers post source code and problem statements that allow the unexpected program behavior they describe to be reproduced.
To start, we conducted an exhaustive web search and gathered as many different locations as possible.
This was done initially by performing web searches using keywords such as ``COBOL forums" to identify the easiest-to-find sites.
Additionally, we manually evaluated each of these sites to determine if they contained links or information on any related or similar sites.
This resulted in an initial collection of forums and chat rooms where users had posted mainframe and COBOL-related technical questions.
Once we exhausted the results from web searches and manual evaluation, the next step was to ask for the recommendations of users and community members from the currently identified locations.
Since the COBOL community predates the web, we felt these additional steps would help capture the harder-to-find locations that may have been forgotten over time.
The questions we posted to these sites asked users of other places that exist in the COBOL community where they would go to get help.
We received multiple responses to each of these posts, and it helped identify missed locations such as mailing lists and chat servers.
This led to an iterative approach in which we asked the same question in every newly found location until we reached a theoretical saturation of community locations.
Overall, we collected more than 25 different web-based community locations focused on mainframe and COBOL.

The next step in this process was to determine how many of these sites actually contain posts related to defective COBOL programs. 
This was done via manual evaluation of the questions and responses asked on each site.
In this process, we followed a set of criteria designed to help identify if the location is used for technical troubleshooting and problem-solving.
This criteria first consisted of asking a simple question about the theme of the site:
Is this a location in which users interact and participate in Q\&A on mainframe technology and COBOL development?

If the answer was yes, the next check was to then attempt and identify at least one "example" post from the location. 
There were three requirements for a post to be considered a valid example:
\begin{itemize}
 \item Does the post ask about a defect solely related to COBOL programming? (i.e. not a discussion of compiler versions, hardware issues, etc).
 \item Does the post contain the needed background information? (i.e. a code snippet, the unexpected output, etc)
 \item Does the post contain replies that point to a solution for the defective program?
\end{itemize}
If at least one example post could be found on the site, the location was considered valid and used for data collection. 
After this filtering, we were left with two primary locations that contained questions with the information needed to reproduce defective COBOL programs. 

The first of these locations selected to be scraped was the IBMMainframe Experts forum~\footnote{\url{https://ibmmainframes.com/forum-1.html}}.
This forum, which has since been abandoned, was seemingly the previously most trafficked forum to seek help with COBOL development. 
It contains over 6,000 posts in the COBOL programming channel, dating back over 15 years.
However, much like the COBOL community as a whole, the older developers that frequented this forum seem to be long gone. 
As such, the amount of questions asked and responses provided has decreased significantly. 

The second location is the tek-tips IT forum by engineering.com~\footnote{\url{https://www.tek-tips.com/threadminder.cfm?pid=209&page=1}}.
This location contains almost 2,500 posts that date back over 20 years.
Since this forum channel is a COBOL programming general discussion, it contains much fewer technical question posts.
Similar to the IBMMainframe, this location is also mostly abandoned, and the frequency of questions has declined significantly over time.

While the goal of this process was to determine the best locations to mine for data, we feel that this collection of public community locations itself serves as a valuable resource to the COBOL community.
Our hope is that the list grows and expands in parallel with work on this topic. 

\subsection{Data Collection}
Once we established the locations, we began the data collection phase.
While manual evaluation of the community locations was feasible due to the limited amount ($<$30), this was not a reasonable option for the posts themselves. 
Overall, there were more than 8,500 different posts spread across our two final collection locations.
This presented a clear need for automation of the collection and preliminary evaluation process.
To do this, a Python-based web-scraper was implemented to collect any relevant posts.
This web-scraper was designed in a similar fashion to our manual evaluation process and had a built-in criterion to determine the initial relevance and reproducibility of the post.
In this case, posts were filtered and selected based on simple search criteria. 
Firstly, posts were identified by a keyword search over the title and post body.
A list of keywords was established containing phrases such as "bug, issue, defect, unexpected" as well as COBOL specific keywords such as  ``\texttt{abend}'' and "\texttt{soc}".
The complete list of keywords used is shown in Table ~\ref{tab:keywords}.

\begin{table}[]
\centering
\caption{List of keywords used to collect posts when conducting the forum web scraping.}
\begin{tabular}{@{}ll@{}}
\toprule
\multicolumn{2}{c}{\textbf{Search Keywords}} \\ \midrule
error & crash \\
unexpected & fail \\
defect & issue \\
broken & soc \\
abend & failed \\
bug & problem \\ \bottomrule
\end{tabular}
\label{tab:keywords}
\end{table}

Posts flagged as containing a keyword were then passed to the second stage of filtering to determine if the post contained any source code.
If both of these criteria were met, the final stage of filtering determined if the post had replies or responses possibly providing an identification of the error and a proposed solution.
This process resulted in a list of 392 posts for manual review and classification.

\subsection{Post Classification}
To evaluate the collected posts, we conducted two separate rounds of classification.
The first round determined which of the 392 collected posts were viable for recreation.
This was done by reviewing the post and assigning both a reproducibility and relevance score.
The second round was designed to identify the type of defect present in posts that were determined reproducible and relevant. 
Together these two types of coding provided a filtering and selection method, as well as additional information on the defect identified in each post.
The details of each round of classification are presented here.

\subsubsection{Reproducibility and Relevance}
To determine how many of these automatically selected posts were actually reproducible and relevant, a manual evaluation and coding process was used.
This type of qualitative coding gave us the opportunity to gauge the reproducibility of the post based on the information presented, as well as the relevance of the type of defect presented.
To ensure consistency in reproducibility and relevance coding, the authors used two independent coding sessions, followed by Cohen's Kappa calculations and discussion~\cite{hsu2003interrater}. 
In the first round of independent coding 10\% of the 392 posts were randomly selected and evaluated for reproducibility and relevance by each author.
Once each author had assigned codes, a Cohen's Kappa score of 0.32 was calculated, which is considered a ``fair" strength of agreement.
After discussion to resolve disagreements, the second round of coding on another 10\% of the data was completed.
For the second round, a score of 0.67 was calculated, indicating a ``good" strength of agreement.
One final round of coding on another 10\% of the data was completed and a final Cohen's Kappa score was calculated at 0.92 which is considered a ``very good" strength of agreement.~\cite{hsu2003interrater}
The remaining ~70\% of posts were then coded for reproducibility and relevance by a single author.

We based the reproducibility on the background information given, the source code posted, and whether or not a resolution was present.
The presence of these three things indicated the post was highly reproducible.
The relevance of the post was based on the expected cause of the defect.
Since the goal of the defective program suite was to capture the most common bugs and issues encountered when developing in COBOL, we determined the relevance based on which part of the mainframe development process was performing unexpectedly.
If the cause of the issue was directly related to the compiler version/options, Job Control Language(JCL) ordering, or system resources, then it was determined irrelevant. 
This reasoning results from the fact that an issue such as a missing compiler flag is not something that can be determined by analysis of the COBOL code alone.
The reproducibility and relevance coding process resulted in a total of 43 posts being recreated for the benchmark suite.

\subsubsection{Orthogonal Defect Classification}
Each of these final posts selected for recreation was then assigned a defect type and defect qualifier as defined orthogonal defect classification based on the definitions presented in \cite{chillarege1992orthogonal}.

All of the final 43 posts were coded independently by the authors, and after discussing and resolving differences, full agreement on the ODC coding was reached.
We chose to use IBM's ODC rather than our own set of defect types since it is an established and well-defined reference point for defect analysis.
For the ODC coding, we focused on just the ODC type and qualifier since these two aspects best fit our purpose of having a lightweight defect classification.
The full list of ODC types and qualifiers used can be seen in Table~\ref{tab:odc}.
This classification provided an additional analysis method and also gives a more detailed breakdown of the type of defective programs contained in the benchmark.

\begin{table}[ht]
\centering
\caption{IBM's Orthogonal Defect Types and Qualifiers defined in \cite{chillarege1992orthogonal}. This is the full list that was used to classify the defect type and qualifier.}
\begin{tabular}{@{}ll@{}}
\toprule
\textbf{ODC Type} & \textbf{ODC Qualifier} \\ \midrule
\textit{Assignment/Initialization} & \textit{Missing} \\
\textit{Checking} & \textit{Incorrect} \\
\textit{Algorithm/Method} & \textit{Extraneous} \\
\textit{Function/Class/Object} & \textit{} \\
\textit{Timing/Serialization} & \textit{} \\
\textit{Interface/O-O Messages} & \textit{} \\
\textit{Relation} & \textit{} \\ \bottomrule
\end{tabular}
\label{tab:odc}
\end{table}

\subsection{Program Recreation}
One of the authors is an experienced developer with over 60 years of professional COBOL development experience who acted as a developer to ensure accurate recreation of the identified defect posts. 
The developer was provided the post describing the defect and instructed to recreate both the defective program as well a fixed version.
The COBOL developer further analyzed the defects present and wrote a problem statement that explains the specific COBOL-based error in each post.
Based on this problem statement, a COBOL program was written containing code that replicated the error.
Finally, the COBOL developer wrote code that solves the defect, to allow the program to execute without issues. 
This was then combined with the defective program so that it contains both the correct and incorrect code.
Next, any accompanying data like job control language files and data files were written and added to the set of programs to provide the needed framework to run each COBOL program.
All of this was organized into a single repository to allow for easy distribution and use.
All of this information was then combined with the ODC type and qualifier previously assigned to provide as much metadata as possible for the COBOL programs.
To ensure quality and minimize individual bias, we also had these programs independently reviewed by an external experienced developer with over 30 years of professional COBOL development experience.
\section{Results}
The work to create the final benchmark suite resulted in contributions in addition to the benchmark suite. 
We present each contribution and result of our work in more detail here.

\subsection{Community Locations}
During our search for sites containing technical questions related to COBOL and mainframe technology we found and analyzed a number of other community hubs.
Locations range from chat-based servers such as discord and slack channels, to official IBM blogs, webcasts, and documentation.
Our manual evaluation of these locations allowed us to categorize and organize them into an additional resource.
Since there is a lack of a COBOL presence in more modern locations, like Stack Overflow, new developers may find it difficult to know where to seek help outside of their colleagues.
This issue is made even worse by the possibility of their colleagues being no more experienced than themselves, as more veterans retire from the workforce.
We hope the central location of all of these resources will provide a starting point for new developers searching the web for help with their COBOL and mainframe questions.
The most up-to-date list of the forums is available on our project web~\cite{projectWeb}.

\subsection{Post Classification}
Once we had selected the forums to mine from the list of community locations, the next step was identifying and collecting relevant posts.
As defined in the methodology, this was first automated, then finalized with manual evaluation.
With the list of 392 posts for manual evaluation, we identify 43 to recreate for our benchmark suite.
Here we present the breakdown of both the original 392 and the orthogonal defect classification coding from the final 43.

\subsubsection{Reproducibility \& Relevance}
To determine the reproducibility and relevance of each post, they were coded with one of three options.
For reproducibility, we used either `yes', `no', or `maybe'.
`Yes' scores were given to posts that contained adequate background information, code snippets, and replies providing a solution.
`Maybe' scores were given when at least two were present, but it couldn't be fully verified they were reproducible.
`No' scores meant the post was lacking at least one major part needed for recreation and therefore was not reproducible.

\begin{figure}
 \centering
 \includegraphics[width=0.8\linewidth]{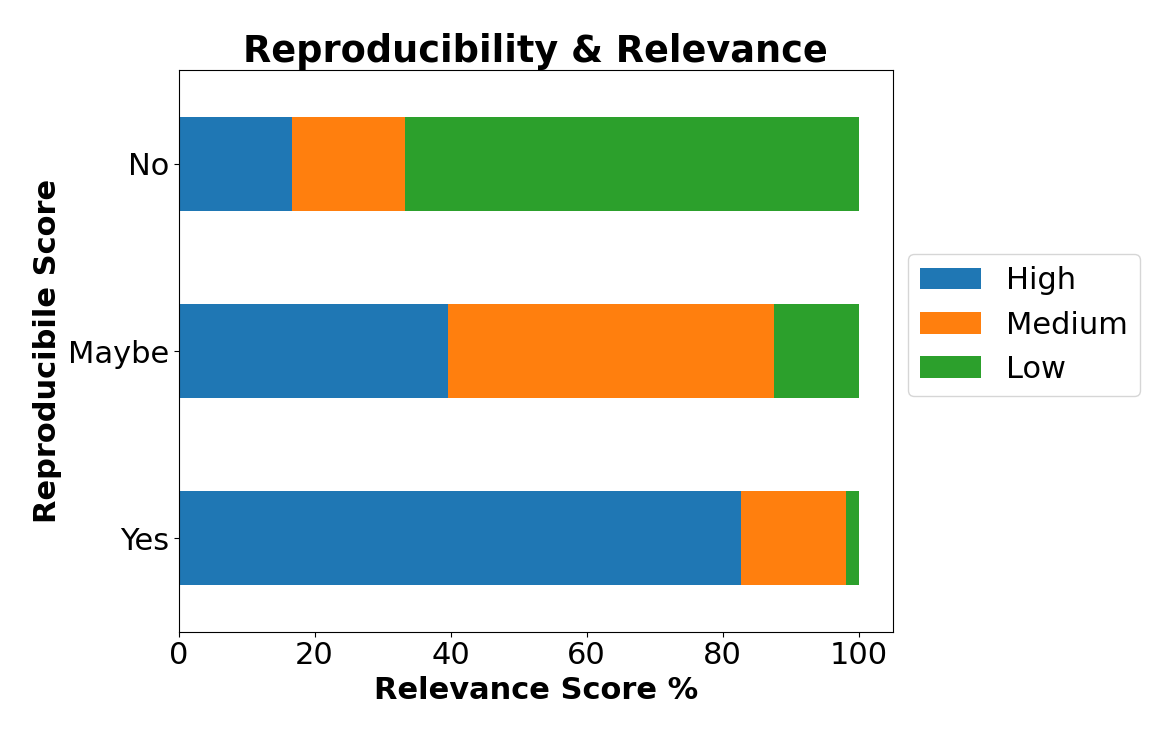}
 \caption{A comparison of the distribution reproducibility and relevance scores. It illustrates a positive correlation between program reproducibility and defect relevance.}
 \label{fig:rep_rel}
\end{figure}

Next, the post was given a relevance score which assigned a priority to the most reproducible posts.
This was a score of either `high', `medium', or `low'.
Posts given a high relevance score were determined to have a defect directly related to mistakes made while writing COBOL code.
Medium posts had the possibility of being directly related to COBOL, but it could not be fully confirmed.
Low scores meant that the defect in the post was not related to COBOL programming but rather caused by outside factors such as a missing compiler flag, or hardware/system resources issues.
The breakdown of reproducibility and relevance scores can be seen in \CalloutTable{rep-rel-tbl}, and that data was also graphed in \CalloutFigure{fig:rep_rel}.

\begin{table}[ht]
\centering
\caption{Reproducibility \& Relevance}
\begin{tabular}{@{}lllll@{}}
\toprule
\multirow{2}{*}{\textbf{Reproducible}} & \textbf{Relevance} & & & \multirow{2}{*}{\textbf{Total}} \\ \cmidrule(lr){2-4}
 & \textit{High} & \textit{Medium} & \textit{Low} & \\ \midrule
\textit{Yes} & 43 & 8 & 1 & 52 \\
\textit{Maybe} & 19 & 23 & 6 & 48 \\
\textit{No} & 4 & 4 & 88 & 96 \\
\textbf{Total} & 66 & 35 & 95 & \textbf{392} \\ \bottomrule
\label{rep-rel-tbl}
\end{tabular}
\end{table}

First, we removed posts that were determined to not contain any recognizable defect.
This filtered out 107 posts from our initial list and left 248 posts containing some type of defect.
Analysis of this data shows that the posts rated as highly reproducible were also the most likely to be coded as highly relevant.
In total, posts coded \emph{Yes} for reproducibility were also coded \emph{High} for relevance roughly 83\% of the time, \emph{Medium} roughly 16\% of the time, and only a single post was coded both \emph{Yes} and \emph{Low}, making up less than 1\%. 
Inversely, posts coded as \emph{No} for reproducible, were also coded as \emph{Low} relevance 67\% of the time, and \emph{Medium} or \emph{High} 17\% of the time.
Finally, posts coded as \emph{No} were more evenly split with \emph{Medium} at 48\%, \emph{High} at 40\%, and \emph{Low} at 12\%.
There are a few explanations for this distribution, as the most reproducible posts contained the most information to evaluate the relevance, resulting in the majority of reproducible posts being considered highly relevant since the cause of the defect was easy to determine.
However, posts with little information scored \emph{No} for Reproducible, were most likely to be rated \emph{Low} since the source of the defect could not clearly be determined.
This reasoning follows with the \emph{Maybe} score, as the relevance rating is more balanced and more likely to be considered \emph{Medium} since moderate, but not complete information is provided.

\begin{figure}
 \centering
 \includegraphics[width=0.9\linewidth]{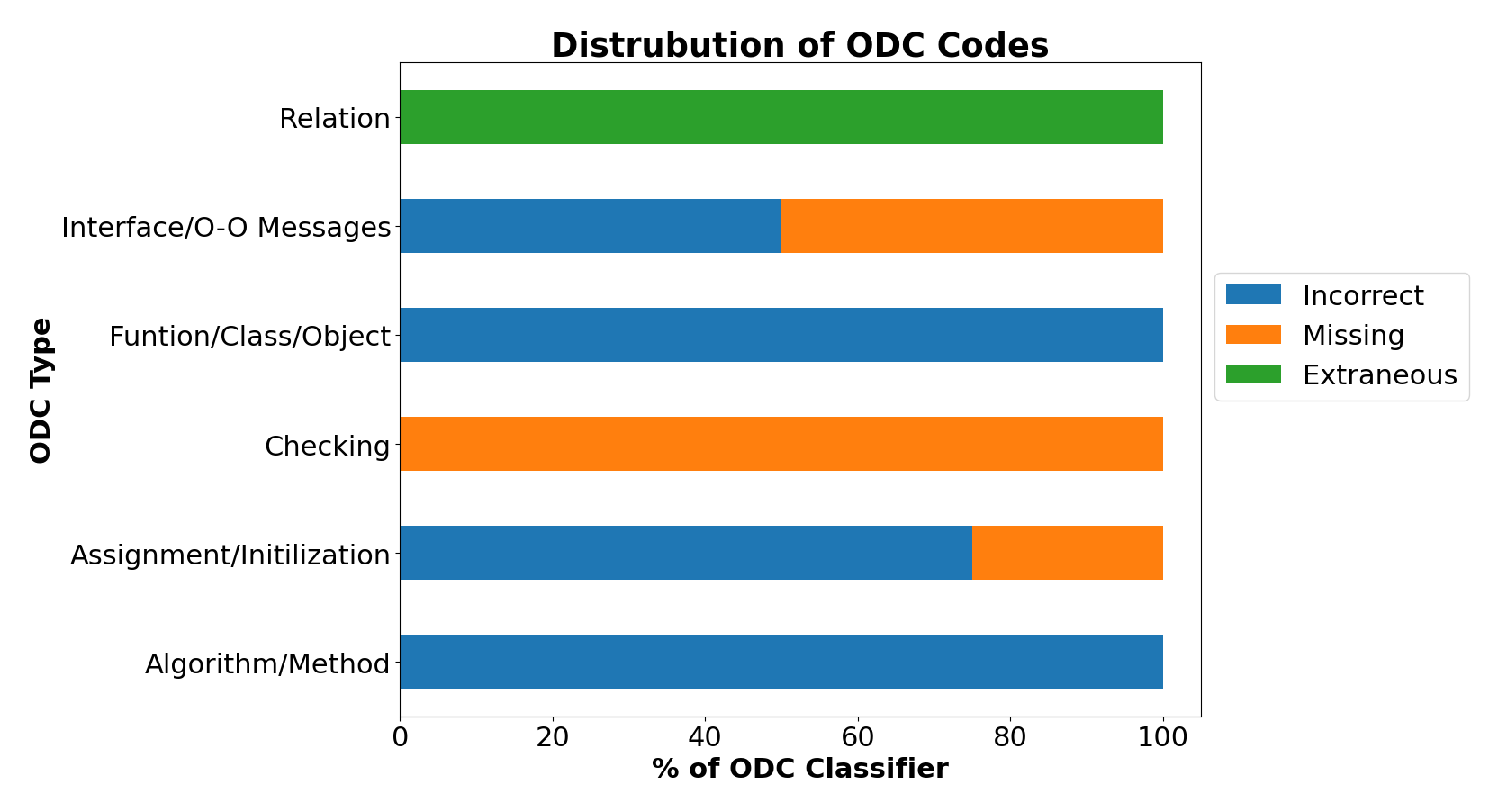}
 \caption{This graph compares the distribution of orthogonal defect types to qualifiers. It illustrates how the vast majority of defect qualifiers were \emph{Incorrect} or \emph{Missing}.}
 \label{fig:odc_class}
\end{figure}

\subsubsection{ODC Code}
\begin{figure}
 \centering
 \includegraphics[width=0.6\linewidth]{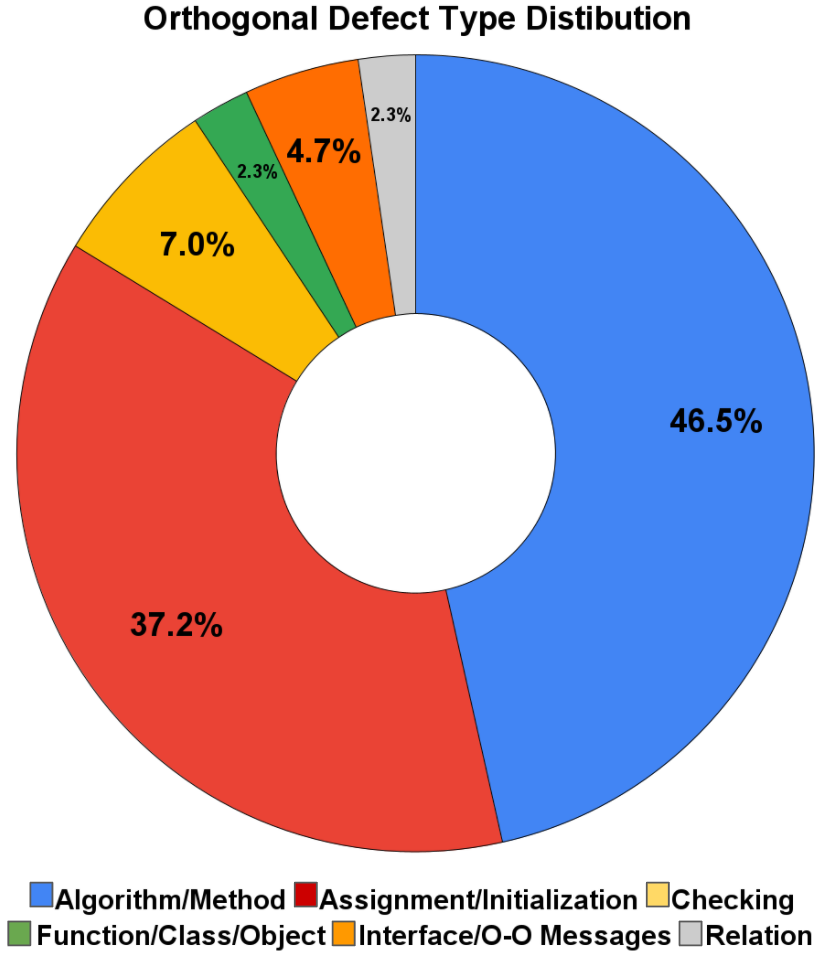}
 \caption{Percentage breakdown of the orthogonal defect type distribution for the final 43 posts selected to be recreated in the benchmark suite.}
 \label{fig:odc-types}
\end{figure}

This produced our final list of 43 forum posts.
Each post was assigned an ODC type and qualifier and the breakdown of those scores can be seen in \CalloutTable{odc-table}.
The distribution of scores was also plotted in \CalloutFigure{fig:odc-types}.
The most common orthogonal defect code type was \emph{Algorithm/Method}, which made up 46.5\% of posts.
Next, the \emph{Assignment/Initialization} type was the second most common at 37.2\%.
The remaining types made up a much smaller portion of the posts, with \emph{Checking} at 7\%, \emph{Interface/O-O Messages} at 4.7\%, and the \emph{Relation} and \emph{Function/Class/Object} both at 2.3\%.

Each post was also assigned an orthogonal defect classification qualifier. 
The distribution of these qualifiers can be seen in Figure \CalloutFigure{fig:odc_class}.
The most common ODC type, \emph{Algorithm/Method}, was assigned the \emph{Incorrect} qualifier 100\% of the time.
The second most common type, \emph{Assignment/Initialization}, had a slight bit more variety with 75\% falling into the \emph{Incorrect} qualifier, and 25\% considered \emph{Missing}.
The following two, \emph{Checking} and \emph{Function/Class/Object}, were both coded 100\% as \emph{Missing} and \emph{Incorrect}, respectively.
Posts with ODC type \emph{Interface/O-O Messages} were split evenly between \emph{Incorrect} and \emph{Missing}, while the single post considered \emph{Relation} was the only one with the \emph{Extraneous} qualifier.

\begin{table}[ht]
\centering
\caption{Orthogonal Defect Classification type and qualifier counts for the final 43 posts selected for recreation.\\}
\begin{tabular}{@{}lllll@{}}
\toprule
\textbf{ODC Type} & \multicolumn{3}{l}{\textbf{ODC Qualifier}} & \textbf{Total} \\ \cmidrule(lr){2-4}
 & \textit{Incorrect} & \textit{Missing} & \textit{Extraneous} & \\ \midrule
\textit{Algorithm/Method} & 20 & 0 & 0 & 20 \\
\textit{Assignment/Initialization} & 12 & 4 & 0 & 16 \\
\textit{Checking} & 0 & 3 & 0 & 3 \\
\textit{Function/Class/Object} & 1 & 0 & 0 & 1 \\
\textit{Interface/O-O Messages} & 1 & 1 & 0 & 2 \\
\textit{Relation} & 0 & 0 & 1 & 1 \\
\textbf{Total} & 34 & 8 & 1 & \textbf{43} \\ \bottomrule
\label{odc-table}
\end{tabular}
\end{table}

\subsection{Benchmark Suite}
Each program contained in the benchmark is designed to be fully functional and is accompanied by all necessary supporting files.
These supporting files include example database tables since some defects were related to database query errors and other example input data.
A job control language or JCL file was also included for each program, to ensure all parts needed for execution were present.

\begin{figure}
 \centering
 \includegraphics[width=0.8\linewidth]{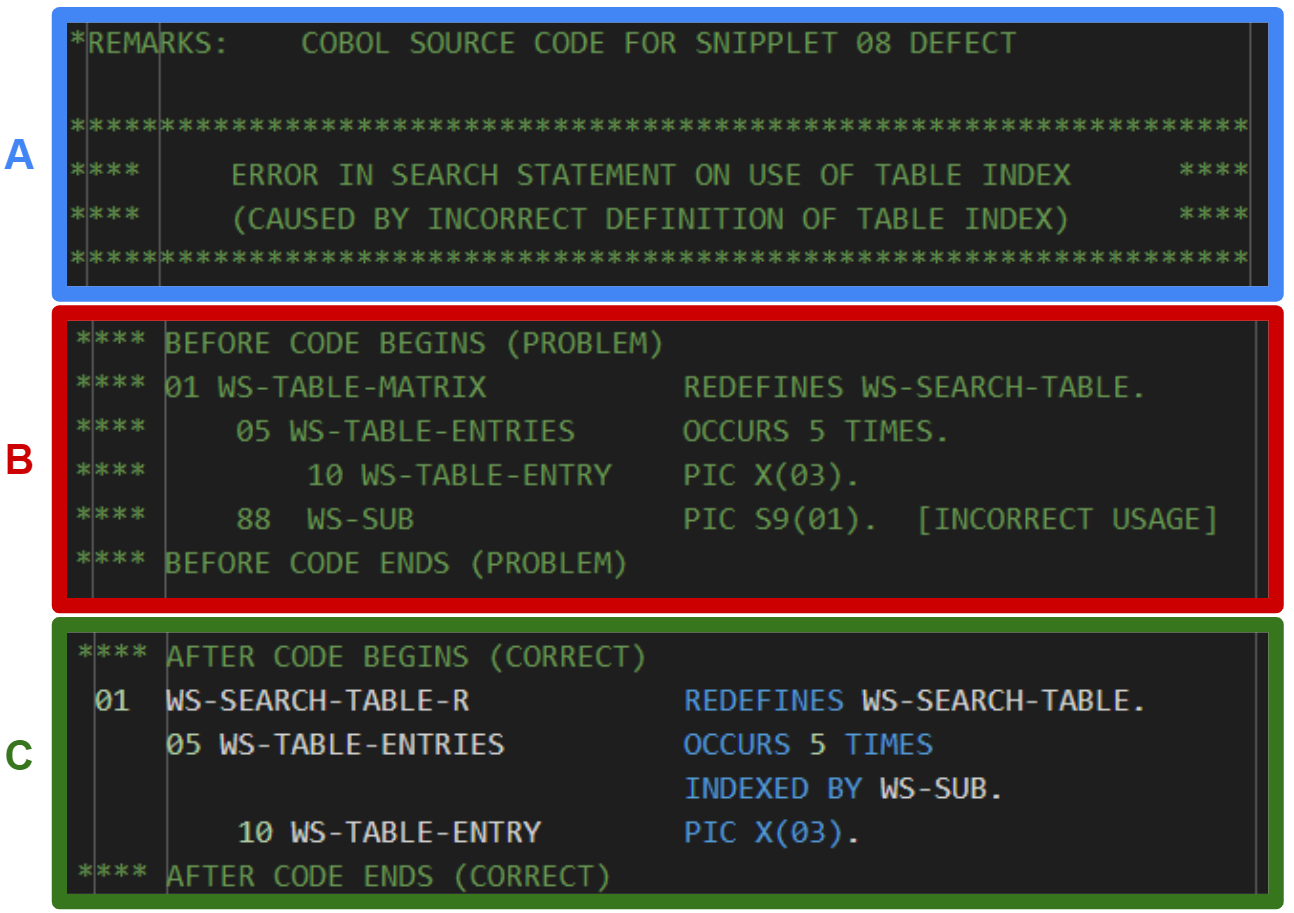}
 \caption{These are snippets of the three most relevant sections of a recreated COBOL program in the benchmark. Part A contains the problem statement, Part B contains the section of defective code, and Part C is the fixed, non-defective code.}
 \label{fig:code_example}
\end{figure}

\CalloutFigure{fig:code_example} shows three of the most relevant parts that make up each program in the benchmark.
Firstly, \CalloutFigure{fig:code_example}-A shows a file header containing the defect number, which maps to the post in the final list of defective forum posts.
Also in the header, is the problem statement as written by the author responsible for recreating the defect.
This problem statement provides yet another identification of the defect presented in the program.
In this example, the problem statement describes that the error was present in a search statement using a table index, and further describes that this error was directly caused by an incorrect definition of a table index.
This matches the ODC type assigned to the defect presented in the post, \emph{Assignment/Initialization} and allows for verification of ODC types and qualifiers.

Next, \CalloutFigure{fig:code_example}-B contains the recreated defective code that attempts to encompass the defect presented in the corresponding forum post.
In this example, the code incorrectly declares a table index, and the specific line with the error is identified.
This code is commented out by default to ensure the program is runnable without modification. 

Finally, \CalloutFigure{fig:code_example}-C presents a solution to the defect that correctly accomplishes the original intent of the program.
Again, comments clearly point out that this is the solution code, and allows for easy comparison between the defective and fixed code.
Since the defect in this program was due to the incorrect definition of a table index, the correct definition is provided while still preserving the other related code.
Additionally, since the defect in \CalloutFigure{fig:code_example} requires an input table, the needed supporting files are included in the benchmark.

\begin{figure}
 \centering
 \includegraphics[width=0.8\linewidth]{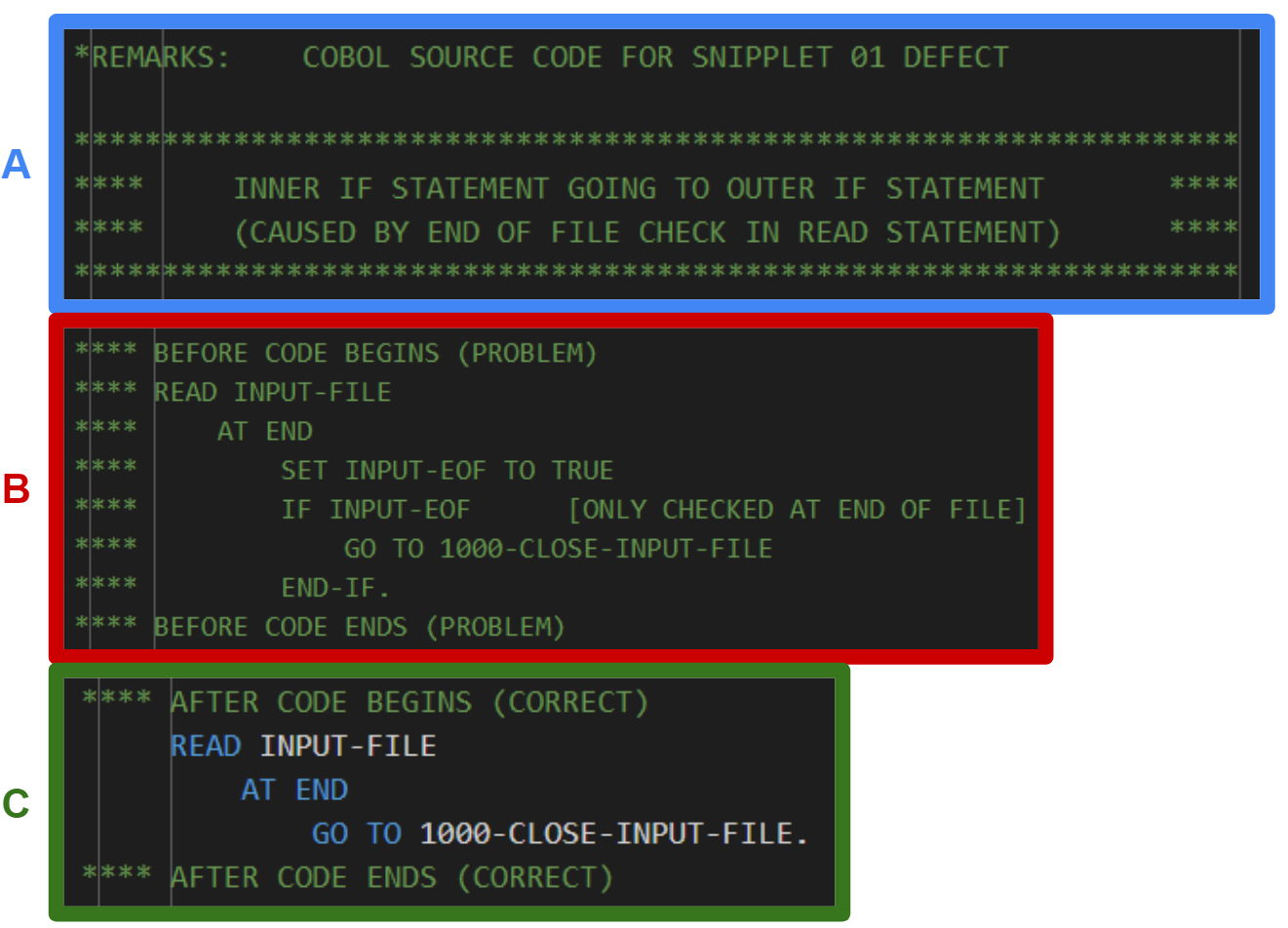}
 \caption{Another example program from the benchmark showing the three most relevant sections. Part A contains the problem statement, Part B contains the section of defective code, and Part C is the fixed, non-defective code.}
 \label{fig:code_example2}
\end{figure}

Another example of a program from the benchmark can be seen in \CalloutFigure{fig:code_example2}.
This example shows the second most common ODC type of \emph{Algorithm/Method}, by demonstrating incorrect logic when reading from a file.
The defect as described by the COBOL developer can be seen in \CalloutFigure{fig:code_example2}-A, where the logic error and the cause of the logic error are defined.
In this case, there is an incorrect check for the end of the file during a read statement, which caused an unexpected jump to an outer if statement.

\CalloutFigure{fig:code_example2}-B shows the incorrect if and read statement logic, and the specific line checking for EOF is identified.
This section also mentions when the end of file check is actually done, giving more insight into why it is incorrect.

Lastly, \CalloutFigure{fig:code_example2}-C shows the simple solution to the defective code, providing an example of how to properly read from a file.
Again, since this defect requires an input file, the needed supporting file is included in the benchmark.

The examples chosen for \CalloutFigure{fig:code_example} and \CalloutFigure{fig:code_example2} give an idea of what a defective COBOL program would look like that capture the two most common ODC types \emph{Assignment/Initialization} and \emph{Algorithm/Method} respectively.

This all provides a multitude of ways this suite could assist COBOL and mainframe tool-smiths and provides validation methods spanning multiple different angles.
To our knowledge, this is the first open-source COBOL resource of this type. 
We hope this suite gorws over time with community contributions and serves as a resource for the mainframe toolsmiths and researchers to further innovate in this space to usher in a new generation of mainframe development.


\section{Discussion}
Here we discuss the results of the community location, post-selection, defect classification, and program recreation process.
We also provide some recommendations for similar work based on the lessons we learned.

\subsection{Location Types}
The types of locations found can be loosely categorized into 5 different groups.
Each group contains multiple different sites, most of which are sponsored by different COBOL-based organizations or companies.

\begin{figure}[!ht]
 \centering
 \includegraphics[width=0.8\linewidth, keepaspectratio]{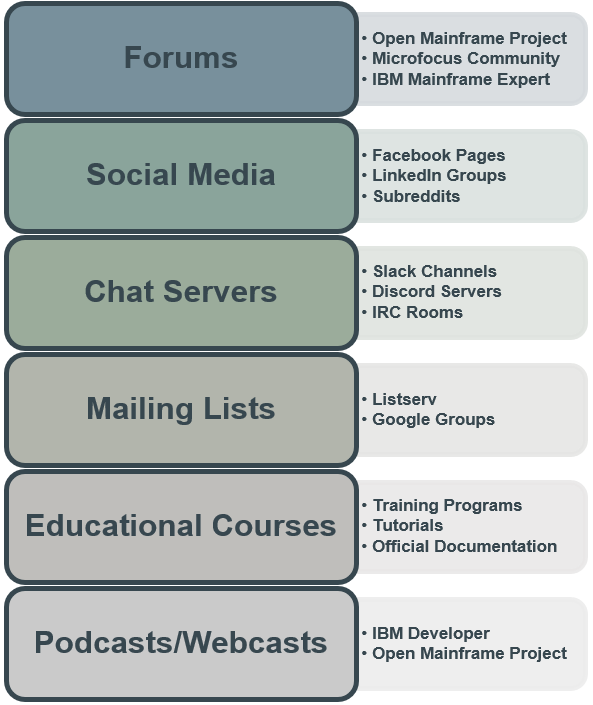}
 \caption{Collection of some of the COBOL/mainframe communities and resources we found throughout our process. They were organized the into different categories based on the location type.}
 \label{fig:cobol_com}
\end{figure}

We identified 5 different COBOL-based forums still in limited use in our search.
The activity on these forums is slowly fading, most likely due to the previously active users leaving the mainframe workforce.
Also, with the ever-expanding number of sites on the internet, more places are filling the gaps left by these old forums.
Since these locations are no longer a likely place to post and receive an answer, the main value they hold is in the vast amount of past posts and data.
This work utilizes a subset of these posts for the purpose of creating a benchmark suite, but the use cases for this are vast.
This can be seen from the sheer number of posts in these locations, some dating back over 20 years and containing tens of thousands of posts.

As these forum locations continue to receive fewer posts and lose active users, newer sites are beginning to fill that role.
Different social media sites that are more popular today are beginning to attract some of the old and new COBOL developers alike.
Places like Reddit, Facebook, and LinkedIn have provided a new outlet for the COBOL community to gather and socialize.
While the purpose of some of these sites, Facebook and LinkedIn in particular, is more suited to allow the professional gathering of large groups and announcements, other locations are beginning to facilitate technical questions.
Places like the mainframe and COBOL subreddits allow for a more informal Q\&A or technical discussion among developers.

Even more so than the social media outlets, different chat servers have seemingly become the most common new place to seek help with COBOL.
This is not a new type of location for COBOL, as we identified multiple IRC groups focused on COBOL.
However, newer chat-type locations have emerged.
Two main locations identified were the Mainframe Enthusiast Discord server, and the Open Mainframe Project sponsored slack channels.
This slack channel has a general chat and a chat specific to their COBOL training program.
This provides a location for both new developers to pose questions, as well as more experienced and complex technical questions. 
These quick chat and fast response locations have begun to provide the same benefits to new developers as the forums did 20 years ago.

In addition to the older forum-style sites and newer chat rooms, we found multiple different COBOL and mainframe-based mailing lists.
Some of these mailing lists dated back decades, and a major site identified contained 100's of different listservs.
While the majority of these listservs are no longer active and past history can only be viewed on an archived site, new locations are stepping in to keep these types of communication channels in place.
Places like Google groups centered around COBOL were identified and still have a large number of users.
The Google groups allow not only provide announcements and updates on mainframe technology, but also another outlet to post technical questions and receive responses.

Outside of technical question locations, we also identified places that help developers keep up to date with new mainframe technology.
One group of these locations are the podcast and webcast-based outlets.
For this work, we only include officially sponsored podcasts or webcasts but acknowledge the many personal, unsponsored casts available.
These type of outlets are yet another step that helps integrate COBOL communities with more modern-day information sources.

We also identified multiple locations that we considered educational in nature.
These types of locations provide free and open-source tutorials or programs to help people learn mainframe technology and COBOL development.
These programs are an essential resource for those wishing to pursue a career with no prior COBOL experience.
These programs have begun to pave the way to get new developers ready and keep current developers up to date with changes in technology.

\begin{tcolorbox}
\small{The retirement of veteran COBOL developers also affects the developer communities they were part of. We noticed the activity in most of the traditionally frequented public communities is slowly fading. 
Thus more effort is needed either to rekindle the activity in such communities or connect them to more contemporary venues to prevent the fading of the treasure-trove of existing mainframe knowledge.}
\end{tcolorbox}

\subsection{Defect Classification}
Analyzing the breakdown of reproducibility and relevance scores, as well as the ODC types and qualifiers assigned to the final posts selected, gives some interesting insight into the type of defects that were recreated.

\subsubsection{Reproducibility and Relevance} 
Firstly, the distribution of reproducibility and relevance scores can provide insight into the reasoning and criteria used for these codes.
Based on the data presented in \ref{fig:rep_rel} there is a clear positive correlation between the reproducibility score of the post and the relevance score.
It can be seen that posts that were considered more reproducible were also considered more relevant.
This is very likely the result of highly reproducible posts containing much more of the information needed to determine the source of the defect.
Since highly reproducible posts contained a large amount of the information needed, that allowed for a more accurate and clear understanding of the defect the post contained. 
With this improved understanding of the defect, we could consider it highly relevant more confidently.
This reasoning also matches the distribution of posts coded ``maybe'' in reproducibility, as the relevance of these was more likely to be categorized as either medium or high.
Again, this logic follows with the analysis of the distribution of posts considered non-reproducible.
Inversely to highly reproducible posts that have adequate information to accurately identify the defect, posts coded as non-reproducible were the most difficult ones to identify the defect in.
Therefore, they were more likely to be considered low relevance, since the type or source of the defect was not able to be identified.

\subsubsection{Orthogonal Defect Classification}
Since orthogonal defect classification is designed to provide an in-process measurement of software defects, breaking down the distribution of ODC types and qualifiers helps to understand the scope of the defects contained in our benchmark suite.
One of the more obvious takeaways from the type distribution in \ref{fig:odc-types}, is that the large majority of defects captured fall into either the ``Assignment/Initialization'' or ``Algorithm/Method'' category.
Our initial hypothesis was that this indicates most of the defects captured are a product of novice developers. 
While they are still representative of common defects created by new developers when learning COBOL, such as incorrectly initializing a variable, or making an incorrect comparison in a loop condition, they may not be fully representative of the type of defects present in industry settings.
This hypothesis was further strengthened in discussions with the COBOL developer recreating the defects, where it was mentioned that some of these posts show the person asking the question had a clear lack of understanding.
We believe this lack of understanding is an indication that these questions are mostly posed by those with limited experience in COBOL and mainframe development.
In one aspect this result is beneficial for utilizing this benchmark suite to address the issues faced by new developers.
From this, it could be assumed that building and testing a tool designed to directly address Assignment/Initialization and Algorithm/Method defects would provide the most direct and immediate benefit to new developers.

This is most likely due to the novice misconceptions about how the code posted would behave.
Additionally, the qualifier for this type was most commonly incorrect.
Again, since the posters seemed to be mostly novice, the most common defects were caused by unexpected program behavior as a result of incorrect logic or program flow.
The most commonly posed solution to these posts required a change to the program's logic or structure.

\subsection{Translating Defects to Programs}

The author with 50+ years of COBOL experience that was tasked with the program recreation shared some of his reflection on the errors he encountered.

\begin{small}
\begin{table}[]
\caption{Summary.\\}
\begin{tabular}{lll}

Search Statement=4 & Read Statement=2 & Move Statement=2 \\
IF Statement=1 & Sort Procedure=1 & Compute Stmt=3 \\
Evaluate Statement=1 & String Statement=1 & Unstring String=2 \\
88 Levels=1 & Convert to Decimal=3 & Occurs Statement=1 \\
Picture Definition=3 & Date Conversion=2 & Record Size=2 \\
Address Pointer=1 & Record Not Available=1 & Set Index=1 \\
Compilation Error=2 & Call Statement=2 & Se Indicator=2 \\
Cursor Processing=2 & Rewrite Statement=1 & Display Data=1

\end{tabular}
\label{tab:summary}
\end{table}
\end{small}

While addressing the 41 defect snippets (6 were not addressed due to insufficient information), it
became clear that most of these issues were encountered by ``beginner'' COBOL programmers, as
opposed to ``experienced” COBOL programmers. But to be fair, many experienced programmers today
encountered many of the defect snippets when first introduced to the COBOL language in past years.
The author's experiences over the years provided a quick insight into cause of the issues and the
recommended code changes to correct these issues. Within each of the 41 defect snippets that were
addressed, the author bookmarked within each program the defect code (commented out) and the
correction code.
A total of 41 programs were developed and tested using ``Micro Focus Developer''~\cite{enterpriseDevloper} to confirm the
corrected code resolved each of the 41 defect snippets. There were an additional 8 COBOL programs
developed and tested to generate the data files, SQL tables, and calling programs for testing the
corrected code as required for certain programs.
In summary, the defect snippets are understandable for “beginner” programmers when first introduced
to the COBOL language, but ``experienced'' programmers rarely repeat these types of program defects
previously experienced. The 41 defect snippets addressed by the author are summarized in the Table~\ref{tab:summary}.

\begin{tcolorbox}
\small{
The extracted benchmark is representative of the issues that beginners face while coding in COBOL. We hope in the future this benchmark gets extended by the community effort to cover example defects faced by intermediate and experienced mainframe developers.   
}
\end{tcolorbox}

\subsection{Lessons Learned}
Each step of this process has provided valuable lessons and experience.
We present our experiences here to provide recommendations for future work related to mainframe developers and the community.

A very important aspect of this work is the interaction and feedback we received from our target community.
This began in the search for community locations by posting in found locations asking that locations users for additional community locations they are familiar with.
While the main goal of this was to ensure full saturation of COBOL communities, it also provided us with valuable feedback on our goals and process.
Initially, our posts in some locations received negative feedback and criticism and helped to identify any preconceived notions we may have had.
Additionally, some users provided recommendations on further work or expansions we should consider.
All of this feedback was very useful and gave us an early window into how the community would receive our work.
Because of this, we were able to tweak and alter our approach to improve our methodology.

Similarly to community feedback, we also noticed the importance of encouraging community contributions.
We found encouraging community contributions even more important in this specific domain.
This was due to the sometimes hostile response when dealing with the COBOL community.
Some of the hostile responses we received made assumptions about our intentions and viewed our work as a potential adversary of COBOL.

We believe the hostility was mainly a result of poor phrasing when posting these questions to the community.
Since the idea of mainframe and COBOL becoming obsolete has prompted many to voice their opinions on the antiquated and dying nature of mainframe technology, developers in this domain are reluctant to accept or provide help.
Some community members reacted negatively to the wording ``most common defects in COBOL'' by questioning ``how do we know there are common defects, are you sure what you're doing is research?''. 
This demonstrated the importance of framing our questions properly and doing our best to remove any possible wording indicating preconceived notions.
We used this to improve our questions to the community, by ensuring we provided the context of our work and our interest in providing resources to benefit the COBOL and mainframe domain.

\begin{tcolorbox}
\small{
The phrasing of the message can have a huge impact on community interactions. We learned that using community-specific vocabulary and providing context generally eases such interactions.
}
\end{tcolorbox}

\subsection{Limitations}
The process followed to create this test suite presented some limitations and threats to validity.
This section discusses these and makes recommendations for similar future work.

One limitation of this work is the Streetlight effect. 
The observational bias causes the researcher to primarily look in the easiest places~\cite{cochran1973controlling}.
This was certainly a factor in our work, especially in the process of finding locations to mine. 
While there exists a huge number of locations that data could be collected from, we focused on the forums that contained a large amount of data that was relatively easy to collect and analyze.
This resulted in all of our data being extracted from two different forum sites.
To alleviate this threat we posted in the identified communities asking about additional COBOL communities until we reached a saturation limit. 

Another threat to validity is a result of the researcher's lack of COBOL experience.
While an experienced COBOL developer assisted with the recreation, the post-filtering and ODC coding were done by researchers with limited COBOL experience.
This means some posts that should have been considered relevant and reproducible, were excluded due to a lack of domain knowledge.
Additionally, it means some of the posts determined relevant and reproducible were not able to be recreated by the COBOL developer, since they lacked vital information not recognized by the researchers.

Finally, there is a limitation in the process we followed to recreate each program. 
Since the program is reproduced based only on the information and source code provided in the forum post, it is reasonable to assume some of the recreated programs do not mimic the behavior of the original defective program exactly.
This is mainly a result of the missing context in the post, such as input files, compiler options, hardware configurations, etc.
We attempted to overcome this limitation by only selecting posts containing a substantial amount of the extra information needed to reproduce the defect.
\section{Conclusion}

Despite being perceived as being outdated; mainframe(escpecially COBOL) technology is here to stay. New developers finding their place in this ecosystem lookup for support and innovation from toolsmiths to further reduce the barrier to entry in the mainframe world. To this end, this paper presents our efforts to synthesize a benchmark suite of COBOL programs representing the defects faced by the ``beginner’’ COBOL developers. These programs are publicly available on our project web. We hope this suite serves as a resource for the mainframe toolsmiths and researchers to further innovate in this space to usher in a new generation of mainframe development.
\section{Acknowledgements}
We would like to thank Phase Change for supporting this work.
Special thanks to Dan Acheff for helping out with reproducing the COBOL programs. This material is based upon work supported by the National Science Foundation (NSF) under Grant Nos. 1850027 and 2008408.

\bibliographystyle{IEEEtran}
\bibliography{ms}

\end{document}